\documentclass[aps,pre,twocolumn,superscriptaddress,floatfix,amsmath]{revtex4}
\usepackage{graphicx}
\usepackage{amsmath,amssymb}
\begin{document}
\title{Effective temperature of active complex matter}
\author{Davide Loi}
\affiliation{
European Synchrotron Radiation Facility, BP 220, F-38043 Grenoble 
France}
\author{Stefano Mossa}
\email[]{stefano.mossa@cea.fr}
\affiliation{
UMR 5819 (UJF, CNRS, CEA) CEA, INAC, SPrAM,
17 Rue des Martyrs, 38054 Grenoble Cedex 9, France}
\author{Leticia F. Cugliandolo}
\email[]{leticia@lpthe.jussieu.fr}
\affiliation{
Universit\'e Pierre et Marie Curie -- Paris VI, LPTHE UMR 7589,
4 Place Jussieu,  75252 Paris Cedex 05, France}
\date{\today}
\maketitle
{\bf
We use molecular dynamics simulations to study the dynamics of an ensemble of interacting self-propelled 
semi-flexible polymers in contact with a thermal bath. Our intention is to model complex systems of 
biological interest. We find that an effective temperature allows one to rationalize the out of equilibrium 
dynamics of the system. This parameter is measured in several independent ways -- from 
fluctuation-dissipation relations and by using tracer particles -- and they all yield equivalent 
results. The effective temperature takes a higher value than the temperature of the bath when the 
effect of the motors is not correlated with the structural rearrangements they induce. 
We show how to use this concept to interpret experimental results and suggest possible innovative 
research directions.
}

{\em Introduction.-} 
Active matter entities, be them particles, filaments or other, absorb energy from 
their environment or internal fuel tanks and use it to carry out motion. 
In this ubiquitous type of soft condensed matter, energy is partially transformed 
into mechanical work and partially dissipated in the form of heat~\cite{active-matter-reviews}.
Units interact directly or through disturbances propagated in the medium.  
The effect of motors can be dictated by the state of the unit and/or
its immediate neighborhood and it is not necessarily fixed by an
external field.  Active matter is thus kept in a non-equilibrium
steady state and presents a number of interesting dynamical features,
unusual mechanical properties, very large responses to small
perturbations, and very large fluctuations.  Realizations of active
matter in biology are manifold and exist at different scales.

Surprisingly enough, studies of active matter have been mainly
analytical so far, based on refined calculations using very stylized
models~\cite{bacterial-th,swarms,BenJacob,Nedelec97,Surrey,Kruse,Hatwalne04,marchetti},
computer simulations of continuum models~\cite{Ignacio}, and numerical
studies of relatively simple lattice models~\cite{Vicsek,Chate}.
Molecular dynamics simulations, that have proven to be so helpful to
elucidate the behavior of complex passive systems, have not been much
employed in this field yet.

Quite independently, passive systems with complex out of equilibrium
dynamics have been the focus of much study. These are systems with
competing interactions between their constituents and no external
energy input that in some conditions (such as sufficiently low
temperature or high density) cannot equilibrate with their
environment. Glassy systems are the typical example.  An important
outcome of the theoretical modeling of these systems, later confirmed
with numerical simulations, is the generation of an effective
temperature, $T_{eff}$~\cite{cugliandolo97}. $T_{eff}$ should be
understood as a time-scales dependent parameter which takes a
thermodynamic meaning in systems with slow dynamics
only~\cite{cugliandolo97,ilg06}. Experimental measurements of $T_{eff}$ in different glassy
systems~\cite{Teff-glycerol,Teff-spinglass,Teff-exp-pol,Teff-granular,Makse-granular,Teff-laponite}
have been performed.

In~\cite{loi08} we took a first step in the direction of analyzing $T_{eff}$ in
active matter. We studied the dynamics 
of an ensemble of self-propelled particles, meant to be a reasonable model for 
real but simple active matter such as bacterial colonies, with molecular dynamics 
simulations. Here we get closer to more complex cases and we study a model of 
filamentous semi-flexible polymers~\cite{liverpool06}. 
We analyze its structure and dynamics and we study $T_{eff}$ by using a 
variety of independent measurements. In particular, we demonstrate that 
tracer particle techniques, that have already been exploited in the study of the 
mechanical properties of real biological systems~\cite{bursac05,wong04},  
can also be useful to obtain a 
direct characterization of the out-of-equilibrium state. 
Details of the calculations 
will be given in a forthcoming publication~\cite{loi10}.

{\em The model.-} We consider the model for a passive semi-flexible linear polymers 
(filaments) of Ref.~\cite{miura01}. The optimal values of the parameters have been 
fixed by preliminary calculations~\cite{miura01,loi10}. Polymer chains are 
coarse-grained and each segment is formed by identical beads, with mass $m$ and 
diameter $r_0$. Each bead is in contact with a thermal environment that is 
described by a random noise and a viscous drag. The dynamical evolution at a bath 
temperature $T$ is therefore controlled by a Langevin equation. 
  
The deterministic mechanical conservative force among beads is given by
intramolecular and intermolecular contributions. Connectivity of the chains is assured by 
harmonic springs acting between nearest-neighbor monomers. Chain rigidity is 
controlled by a purely repulsive interaction between next nearest-neighbor monomers. 
The actual choice of parameters allows one to control the flexibility of the 
chains, ranging from semi-rigid to completely flexible~\cite{miura01,loi10}.  

Having in mind that no simple technique can take into account chemical activity 
in a molecular dynamics calculation, physical insight has been used to
choose a reasonable model for motor action. Only a fixed fraction of polymers 
are provided with motors that have a gentle but still 
detectable effect on the behavior of the system.
We localize the non-conservative motor activity on the monomer at the 
center of the polymer~\cite{note1}. 
The motor effect is given by a time series of random isotropic kicks 
generated by a suitable stochastic process~\cite{loi08}. 
The action of the motors is independent of the structural rearrangements they induce
({\em adamant}~\cite{shen04} motors). 
The force of all motors, $f^M$,  is the same
and a fraction of the mean conservative 
mechanical force acting on the equivalent passive system, $\overline F$. 
We thus quantify the motor activity with the control parameter $f=f^M/\overline F$. 
More details on the model are given in 
Refs.~\cite{loi08,loi10}.

We express all quantities in dimensionless reduced units~\cite{allen89}
that we associate to reasonable values of the experimentally measured 
counterparts~\cite{klimov99}. The typical unit of energy is $2 k_B T$; 
typical lengths are measured in units of $0.4$~nm implying forces of order
 $20$~pN at ambient temperature, $T=300$~K. 

\begin{figure}[t]
\vspace{0.15cm}
\includegraphics[width=0.40\textwidth]{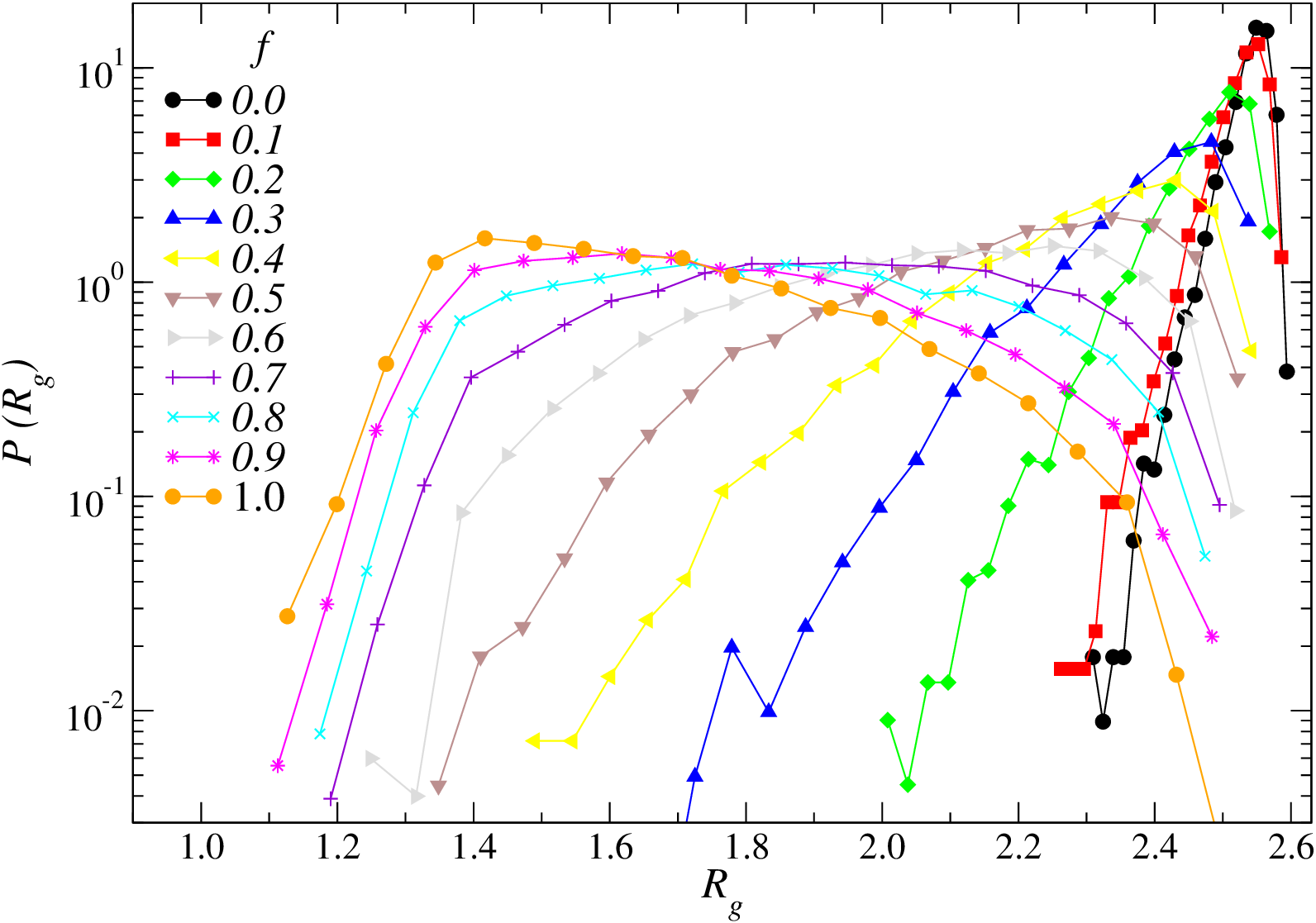}
\caption{(Color online.)
Probability distributions of the gyration radius, $R_g$, at the investigated motor activities, $f$. 
The structure is highly non trivial with an average value shifting towards lower values 
upon increasing $f$. The passive case ($f=0$) is also shown.
}
\label{fig:gyration-vs-nm} 
\end{figure}

{\em Structure.-} We first investigate the structure of the
semi-flexible active polymer melt by keeping all external macroscopic
conditions fixed and varying the motor activity $f$.  The analysis of
the static structure factors and pair distribution functions (not
shown) is consistent with a picture in which the average nearest
neighbor distance between beads pertaining to different polymers
decreases upon increasing $f$. Overall, the motors have the mixed
effect of making the melt more compact and more disordered
simultaneously. This picture is corroborated by the extremely complex
probability distributions of the gyration radius, $R_g$, which are
highly non-Gaussian and very asymmetric
(Fig.~\ref{fig:gyration-vs-nm}).  Motor activity has the effect of
pushing the filaments closer and these, at the same time, fold
substantially. We interpret this behavior in terms of a competition
between two effects: crowding, which is related to excluded volume
effects and tends to slow down the dynamics, and folding of the
chains, which dynamically lifts the topological constraints, therefore
decreasing entanglement. These have important consequences on the
long-time evolution of the system, as it is verified below.
\begin{figure}[t]
\vspace{0.25cm}
\includegraphics[width=0.40\textwidth]{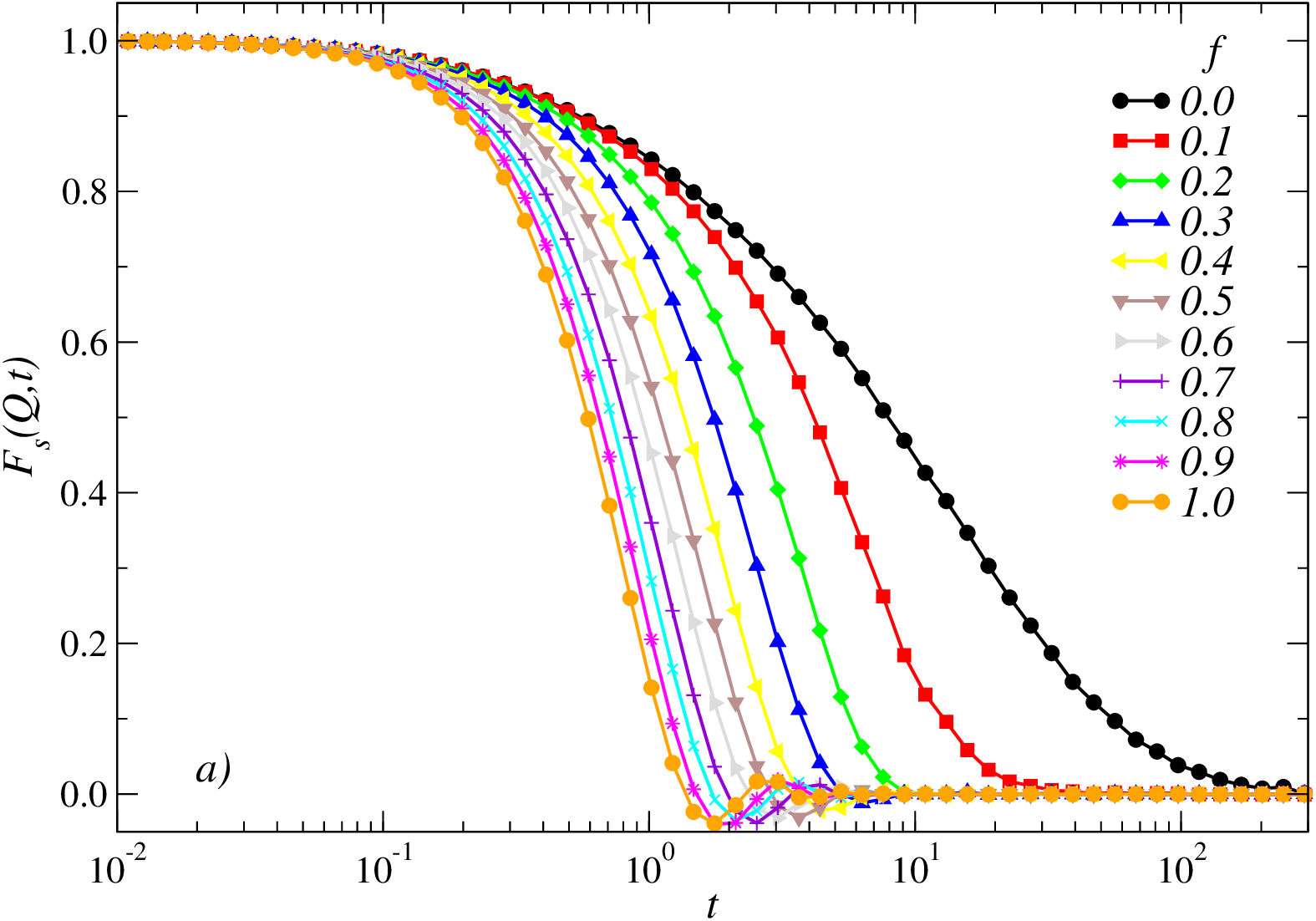}
\vspace{0.25cm}

\hspace{0.2cm}\includegraphics[width=0.38\textwidth]{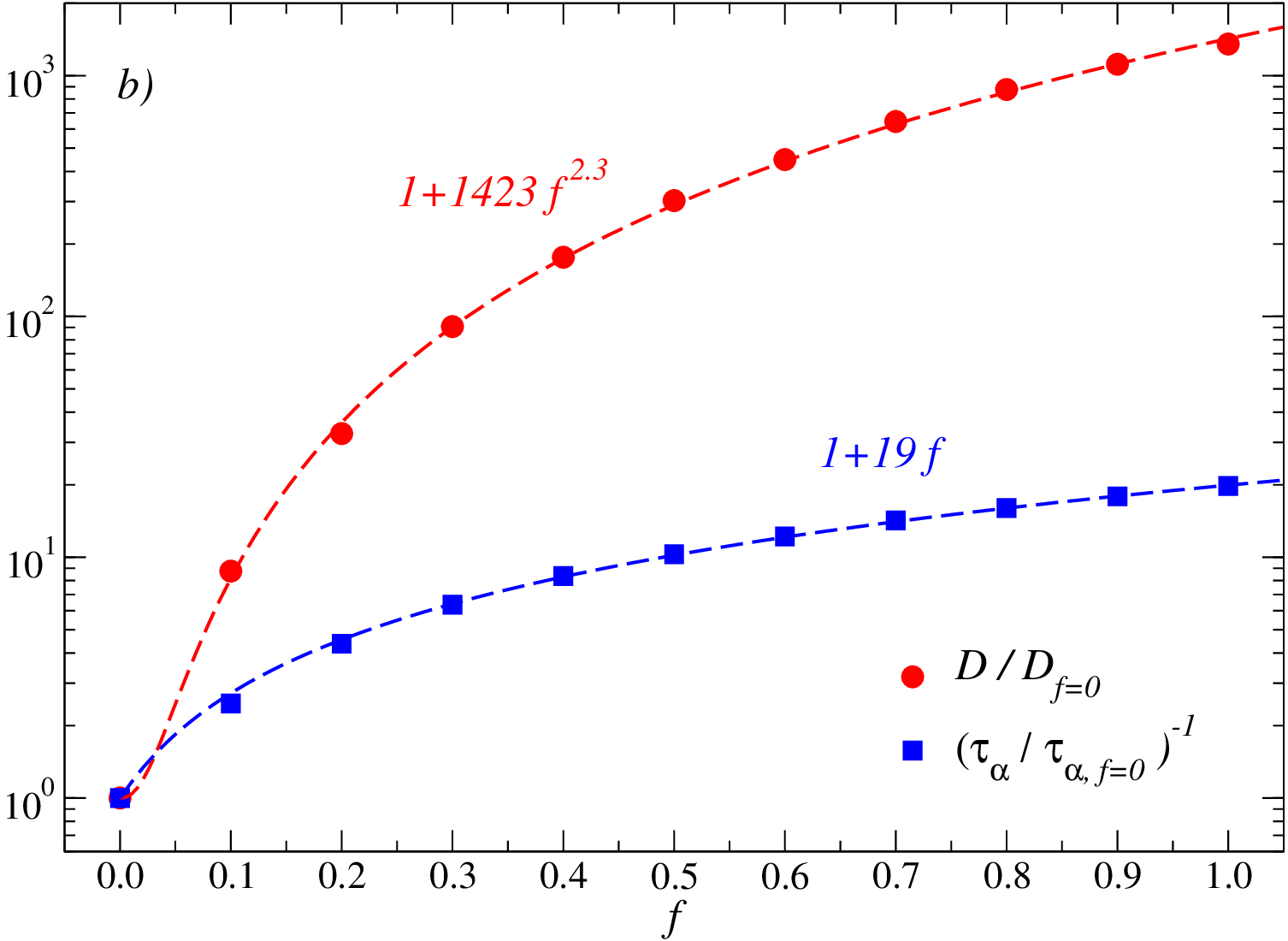}
\caption{(Color online.)
a) Time-dependence of the intermediate scattering function, $F_s(Q,t)$, at the wave-vector 
of the first maximum of the static structure factor (activities increase to the left). 
b) Diffusion coefficient, $D$, and inverse structural (self-)relaxation time, $\tau_\alpha$, 
as a function of $f$. Data are normalized to the value for the passive system and
follow power laws (dashed lines), with the indicated exponents.
}
\label{fig:msd-vs-f} 
\end{figure}

\begin{figure}[t]
\vspace{0.25cm}
\includegraphics[width=0.40\textwidth]{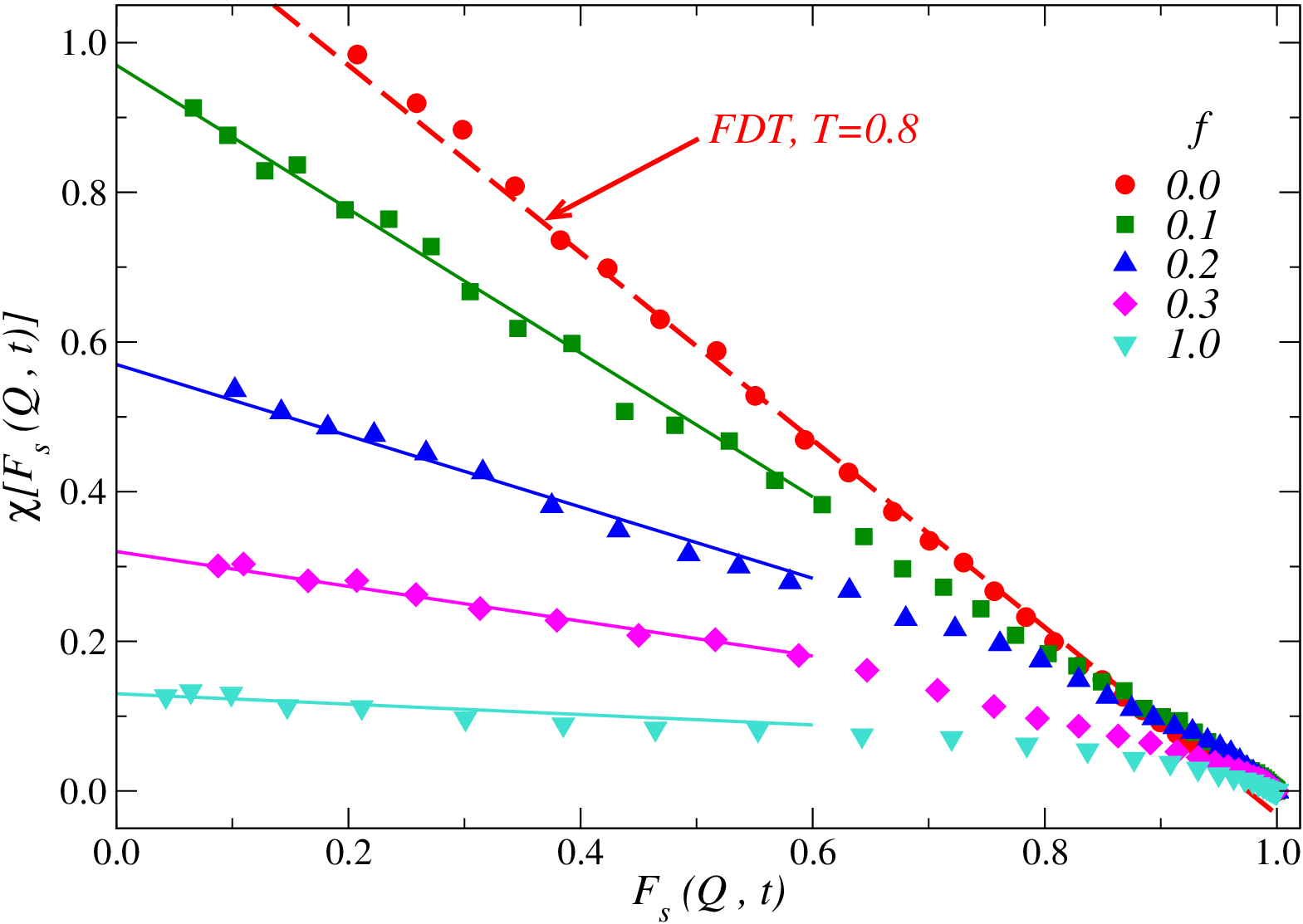}
\caption{(Color online.)
Fluctuation-Dissipation relations at the indicated values of $f$. 
The red data on top (closed circles) are for the passive system and
the dashed (red) line has slope $-1/T$ confirming that FDT holds in this 
case. The solid lines are linear fits used to estimate the value of $T_{eff}$.
}
\label{fig:R-C-vs-f} 
\end{figure}

{\em Dynamics.-}
We next quantify the effect of motor activity on the long time dynamics. 
Figure~\ref{fig:msd-vs-f}~a) shows the self-intermediate scattering function, 
$F_s(Q,t)$, at the wave-vector corresponding to the main peak in the static 
structure factor and as a function of time for the investigated motor forces. 
The mean-squared displacement data (not shown) confirm the picture whereby 
the collective dynamics of the filaments gets faster under stronger motors. 
This is consistent with the gyration radius data and supports our 
interpretation: the augmented folding of the filaments seems to solve local 
topological constraints and diminishes the effect of entanglement. 
A quantitative analysis is presented in Fig.~\ref{fig:msd-vs-f}~b), where we 
plot the dependence of the diffusion coefficient and the relaxation time 
upon the strength of the motors. Quantities are shown renormalized to the value 
of the passive system. It is very interesting to note that the two sets of data 
are consistent with very simple functional forms (dashed lines), 
$D/D_{f=0}\simeq 1+1423\times f^{2.29}$ and $(\tau/\tau_{f=0})^{-1}\simeq 1+
19\times f$, respectively. 

Strikingly, the $f$-dependence of the diffusion coefficient is reminiscent of what 
was found in~\cite{palacci10,howse07}. For the active colloidal 
particles used in these studies, the effective diffusion coefficient can be written 
as $D/D_o=(1+1/9 P_e^2)$, where $D_o$ is the diffusion coefficient of the passive 
system and $P_e$ is the Peclet number, which characterizes the particle 
activity~\cite{palacci10,baskaran09}. In our case the exponent ($2.3\pm 0.1$) is 
close to $2$, the differences being possibly ascribed to the uncertainties on the 
determination of $D$ or to the presence of corrections to the proposed scaling. 
We speculate that here $f$ plays the role of the Peclet number. A similar rationale 
should be valid for the striking linear dependence of the inverse of the 
relaxation time.

{\em Effective temperatures.-}
The analysis above proves that both structure and dynamics of an active system are 
influenced by motor activity in a very complex fashion. We ask now whether it
is possible to embed all this complexity in a single parameter, also prone to be
directly determined in experiments. The chosen quantity is the effective temperature 
$T_{eff}$. One way of measuring the effective temperature in a non-equilibrium
steady state consists in using the generalization of the fluctuation-dissipation 
theorem to out-of-equilibrium conditions~\cite{cugliandolo97}. The most visual way of analyzing
the data is to construct a  parametric 
relation between the integrated linear response, $\chi_{AB}(t)$, 
and the correlation $C_{AB}(t)$ of a wisely chosen pair of observables, $A$ and $B$, 
and associating minus the inverse of its slope with a possibly time-dependent parameter 
$T_{eff}$(t)~\cite{note2,cugliandolo97}. Such measurements were performed in different 
out-of-equilibrium conditions~\cite{glassy-Teff,Berthier-Barrat,Kolton,Makse}
demonstrating that $T_{eff}(t)$ is (asymptotically) piecewise constant taking, typically, two values: 
the one of the environmental bath at short times and an additional one characterizing 
the structural relaxation at long times, in cases with slow dynamics.

\begin{figure}[h]
\vspace{0.25cm}
\includegraphics[width=0.40\textwidth]{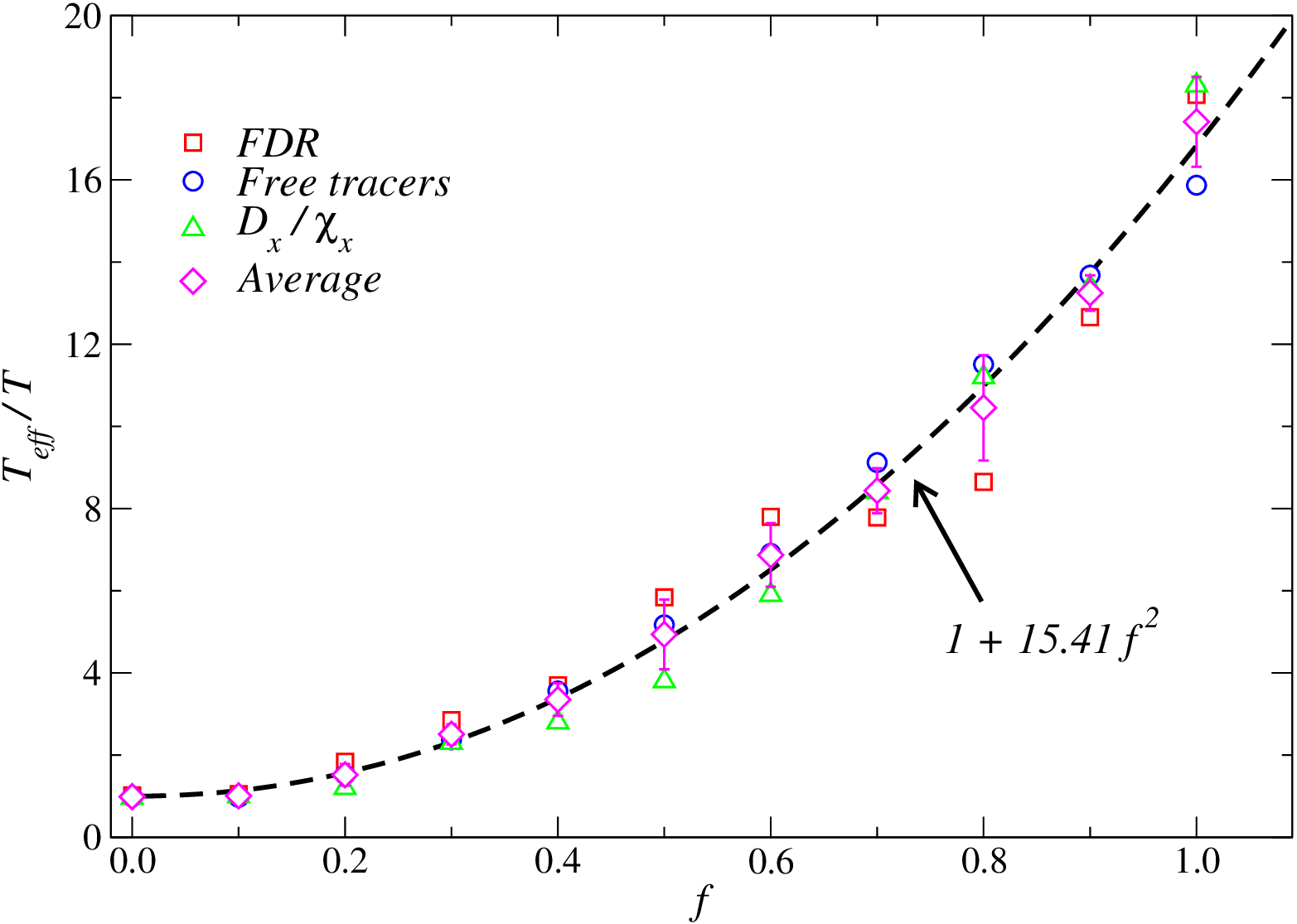}
\caption{(Color online.)
Summary of the entire set of data generated in this study. The power-law dependence of 
the effective temperature on motor activity, $T_{eff}/T=1+\gamma f^2$, is also shown  
(dashed line).
}
\label{fig:Teff-final} 
\end{figure}

We performed these measurements for different values of the forcing
$f$ and the results are shown in Fig.~\ref{fig:R-C-vs-f}. (The needed observables are chosen 
such that the considered correlation functions are the  $F_s(Q,t)$ of Figure~\ref{fig:msd-vs-f}~a). 
Details can be found in Ref.~\cite{Berthier-Barrat}.) 
The curves clearly show two regimes. The
first one, corresponding to high frequencies, represents the fast
vibrations of the monomers, before any structural relaxation takes
place.  The second regime, at long times, describes the actual
structural relaxation in the sample. In all cases, the parametric plot
is rather well described by a straight line, implying that $T_{eff}$ is a constant over this time-scale. 
These data are also shown
in Fig.~\ref{fig:Teff-final} with open (red) squares.

The construction detailed above is very powerful, direct, and easily exploitable in 
computer simulations. In experiments, however, its implementation is far from obvious.
This difficulty has been circumvented by considering the dynamics of 
tracers (or intruders), normally of spherical shape. These objects of micrometric 
size are nowadays quite simple to control and manipulate by active microrheology 
techniques. When immersed in an extended complex environment,
the latter acts as an external reservoir and the intruders' dynamics 
provide a wealth of information on the underlying matrix state. In our 
simulations we have implemented these ideas in two ways.

A possible realization of an adapted thermometer that couples to the
structural relaxation of the sample~\cite{cugliandolo97} is a free
tracer particle with mass much larger than the one of the polymer
beads. Its kinetic energy can be related to the system's effective
temperature via an equipartition theorem~\cite{Berthier-Barrat,Makse}.  Our
results are shown in Fig.~\ref{fig:Teff-final} as (blue) open circles.

Another measurement of $T_{eff}$ is based on the comparison
between (free) diffusion and (driven) mobility of tracer particles
immersed in the active sample. Conforming to the Einstein relation,
$T_{eff}$ should be the parameter linking the tracers' mean-square
displacement and their driven displacement under a weak external
applied force, in the long time limit in which the dynamics is
diffusive.  Data for $T_{eff}$ obtained in this way are included in
Fig.~\ref{fig:Teff-final} as open (green) triangles.

Figure~\ref{fig:Teff-final} collects all our results in the form
$T_{eff}/T$ against $f$. We also include (diamonds) average values
with error bars calculated from the entire set of data. The figure
demonstrates that independent determinations of $T_{eff}$ yield
consistent results and suggests that, in our system, there is a single
frequency-independent $T_{eff}$ parameter, with a precise
thermodynamic meaning.  Data tend to $T_{eff}=T$ in the limit of
vanishing activation, as they should, and the deviation from the
environmental temperature monotonically increases with increasing
forcing. Most importantly, data are represented by the empirical law
$T_{eff}/T=1+\gamma f^2$, with $\gamma=15.41$ (dashed
line). Remarkably, this finding supports our conjecture that the
parameter $f$ plays here a role analogous to the Peclet number for
colloidal active particles considered in a few very recent
experiments~\cite{palacci10,baskaran09}.

In conclusion, we studied the out of equilibrium behavior of a system
of active motorized filaments, a model for complex biological
structures, using a genuine microscopic approach that is still quite
innovative in the field.  We described the influence of motor activity
on both structure and dynamics, finding a rationale for very similar
recent experimental observations. We confirmed the relevance of the
effective temperature, a concept well established in statistical
mechanics of glassy systems, for complex active matter.  Most
important, we demonstrated that this parameter, relatively simple
to measure by microrehology techniques, can be directly related to
chemical activity, a process of extraordinary complexity.  It
would be interesting to explore the possibilities offered by the
molecular dynamics approach used in this paper to clarify the extremely complex
phenomenology of cell mechanical stability~\cite{cyto,mizuno07}. In particular,
one could try to give an answer to the
question as to how mechanical properties (such as elastic moduli) change with
motor activity and whether these changes can be rationalized in terms
of an effective temperature.

\begin{thebibliography}{99}
\bibitem{note1}
We note that, in general, this position does not coincide with the center 
of mass of the filament and it is fixed. We are aware that this situation 
is at variance with the case of real systems, where molecular motors are 
able to slide along the filaments. However, we do not consider this as a 
serious limitation to our approach, since we follow the dynamics of the 
system on limited time scales, over which motor activity can be indeed 
taken as  localized. Also, we average our results over a substantial 
number of instances of  motor realizations, mimicking in an indirect way 
motor motion relative to the filaments.
\end{thebibliography}
\end{document}